\pageno=48
\font\twelverm=cmr12
\font\elfrm=cmr10 scaled 1097
\font\tenrm=cmr10
\font\tentt=cmtt10
\font\tenit=cmti10
\font\tensl=cmsl10
\font\elfsl=cmsl10 scaled 1097
\font\ninerm=cmr9
\font\eightrm=cmr8
\font\sixrm=cmr6
\font\twelvei=cmmi12
\font\twelvebi=cmmib10 scaled\magstep1
\font\tenbi=cmmib10
\font\elfi=cmmi10 scaled 1097
\font\ninei=cmmi9
\font\eighti=cmmi8
\font\sixi=cmmi6
\font\twelvesy=cmsy10 scaled\magstep1
\font\elfsy=cmsy10 scaled 1097
\font\ninesy=cmsy9
\font\eightsy=cmsy8
\font\sixsy=cmsy6
\font\twelvebf=cmbx12
\font\elfbf=cmbx10 scaled 1097
\font\ninebf=cmbx9
\font\eightbf=cmbx8
\font\sixbf=cmbx6
\font\twelvett=cmtt12
\font\ninett=cmtt9
\font\eighttt=cmtt8
\font\twelveit=cmti12
\font\elfit=cmti10 scaled 1097
\font\nineit=cmti9
\font\eightit=cmti8
\font\twelvesl=cmsl12
\font\ninesl=cmsl9
\font\eightsl=cmsl8
\font\twelveex=cmex10 scaled\magstep1
\font\bsi=cmbsy10 scaled\magstep1
\font\si=cmbsy10
\newskip\ttglue
\def\twelvepoint{\def\rm{\fam0\twelverm}%
\def\sl{\fam\slfam\twelvesl}%
\textfont0=\twelverm   \scriptfont0=\ninerm  \scriptscriptfont0=\sevenrm%
\textfont1=\twelvei    \scriptfont1=\ninei   \scriptscriptfont1=\seveni%
\textfont2=\twelvesy   \scriptfont2=\ninesy  \scriptscriptfont2=\sevensy%
\textfont3=\twelveex   \scriptfont3=\twelveex\scriptscriptfont3=\twelveex%
\textfont\itfam=\twelveit    \def\it{\fam\itfam\twelveit}%
\textfont\slfam=\twelvesl    \def\sl{\fam\slfam\twelvesl}%
\textfont\ttfam=\twelvett    \def\tt{\fam\ttfam\twelvett}%
\textfont\bffam=\twelvebf   \scriptfont\bffam=\ninebf%
\scriptscriptfont\bffam=\sevenbf     \def\bf{\fam\bffam\twelvebf}%
\ttglue=.5em plus .25em minus.15em%
\normalbaselineskip=15pt\rm%
\setbox\strutbox=\hbox{\vrule height9.75pt depth4pt width0pt}}
\def\elfpoint{\def\rm{\fam0\elfrm}%
\def\sl{\fam\slfam\elfsl}%
\textfont0=\elfrm\scriptfont0=\ninerm\scriptscriptfont0=\sixrm%
\textfont1=\elfi\scriptfont1=\ninei\scriptscriptfont1=\sixi%
\textfont2=\elfsy\scriptfont2=\ninesy\scriptscriptfont2=\sixsy%
\textfont3=\tenex\scriptfont3=\tenex\scriptscriptfont3=\tenex%
\textfont\itfam=\elfit\def\it{\fam\itfam\elfit}%
\scriptscriptfont\bffam=\sixbf\def\bf{\fam\bffam\elfbf}%
\ttglue=.45em plus.25em minus.15em%
\normalbaselineskip=13.5pt\rm%
\setbox\strutbox=\hbox{\vrule height7.8pt depth3.15pt width0pt}}

\def\tenpoint{\def\rm{\fam0\tenrm}%
\def\sl{\fam\slfam\tensl}%
\def\tt{\fam\ttfam\tentt}%
\textfont0=\tenrm\scriptfont0=\sevenrm\scriptscriptfont0=\fiverm%
\textfont1=\teni\scriptfont1=\seveni\scriptscriptfont1=\fivei%
\textfont2=\tensy\scriptfont2=\sevensy\scriptscriptfont2=\fivesy%
\textfont3=\tenex\scriptfont3=\tenex\scriptscriptfont3=\tenex%
\textfont\itfam=\tenit\def\it{\fam\itfam\tenit}%
\textfont\ttfam=\tentt\def\tt{\fam\ttfam\tentt}%
\scriptscriptfont\bffam=\fivebf\def\bf{\fam\bffam\tenbf}%
\ttglue=.5em plus.25em minus.15em%
\normalbaselineskip=12.5pt\rm%
\setbox\strutbox=\hbox{\vrule height8.5pt depth3.5pt width0pt}}

\def\ninepoint{\def\rm{\fam0\ninerm}
 \textfont0=\ninerm   \scriptfont0=\sixrm \scriptscriptfont0=\fiverm
 \textfont1=\ninei    \scriptfont1=\sixi  \scriptscriptfont1=\fivei
 \textfont2=\ninesy   \scriptfont2=\sixsy \scriptscriptfont2=\fivesy
 \textfont3=\tenex    \scriptfont3=\tenex \scriptscriptfont3=\tenex
 \textfont\itfam=\nineit    \def\it{\fam\itfam\nineit}
 \textfont\slfam=\ninesl    \def\sl{\fam\slfam\ninesl}
 \textfont\ttfam=\ninett    \def\tt{\fam\ttfam\ninett}
 \textfont\bffam=\ninebf   \scriptfont\bffam=\sixbf
  \scriptscriptfont\bffam=\fivebf     \def\bf{\fam\bffam\ninebf}
\ttglue=.5em plus .25em minus.15em
\normalbaselineskip=11pt\rm
\setbox\strutbox=\hbox{\vrule height7.7pt depth3pt width0pt}}

\def\eightpoint{\def\rm{\fam0\eightrm}
 \textfont0=\eightrm  \scriptfont0=\sixrm  \scriptscriptfont0=\fiverm
 \textfont1=\eighti   \scriptfont1=\sixi   \scriptscriptfont1=\fivei
 \textfont2=\eightsy  \scriptfont2=\sixsy  \scriptscriptfont2=\fivesy
 \textfont3=\tenex    \scriptfont3=\tenex  \scriptscriptfont3=\tenex
 \textfont\itfam=\eightit  \def\it{\fam\itfam\eightit}
 \textfont\slfam=\eightsl  \def\sl{\fam\slfam\eightsl}
 \textfont\ttfam=\eighttt  \def\tt{\fam\ttfam\eighttt}
 \textfont\bffam=\eightbf  \scriptfont\bffam=\sixbf
  \scriptscriptfont\bffam=\fivebf  \def\bf{\fam\bffam\eightbf}
\ttglue=.5em plus.25em minus.15em
\normalbaselineskip=9pt\rm
\setbox\strutbox=\hbox{\vrule height7pt depth2pt width0pt}}

\font\rmb=cmr12 scaled \magstep 1
 1
\font\rmc=cmr12 scaled \magstep 2
\font\rmd=cmr12 scaled \magstep 3
\font\rme=cmr12 scaled \magstep 4
 2
\font\ita=cmti12
\font\itb=cmti12 scaled \magstep 1
 3
 4
 5
\font\mitb=cmmi10 scaled \magstep 2
\font\mitc=cmmi10 scaled \magstep 3
\font\mitd=cmmi10 scaled \magstep 4
 5

 1
\def\title#1#2#3{\par\ifnum\prevgraf>0 \npageno=0 \fi
   \global\shortno=0\global\secno=0\global\tabno=0\global
   \figno=0\global\footno=0\global\refno=0\goodbreak\ifnum\npageno=0
   \par\vskip 2cm plus2.5mm minus2.5mm\fi{\noindent\textfont1=\mitd
   \rmd #1$\vphantom{\hbox{\rmd y}}$}\par
   \penalty1000\vskip8pt\noindent{\itb #2$\vphantom{\hbox{\itb y}}$}{\par
   \penalty1000\vskip7.5pt\baselineskip11.5pt\hrule height 0.1pt\par
   \penalty1000\vskip7pt{\tenpoint\noindent\rm #3$\vphantom{y}$}\par
   \penalty1000\vskip6pt\hrule height 0.1pt
   \par\penalty1000\vskip9pt\noindent}}
\def\stitle#1#2#3{\par\ifnum\prevgraf>0 \npageno=0 \fi
   \global\shortno=0\global\secno=0\global\tabno=0\global
   \figno=0\global\footno=0\global\refno=0\goodbreak\ifnum\npageno=0
   \par\vskip 1.6cm plus2mm minus2mm\fi{\noindent\textfont1=\mitc
   \rmc #1$\vphantom{\hbox{\rmc y}}$}\par
   \penalty1000\vskip6.5pt\noindent{\ita #2$\vphantom{\hbox{\itb y}}$}{\par
   \penalty1000\vskip6pt\baselineskip10pt\hrule height 0.1pt\par
   \penalty1000\vskip5.75pt{\ninepoint\noindent\rm #3$\vphantom{y}$}\par
   \penalty1000\vskip4.8pt\hrule height 0.1pt
   \par\penalty1000\vskip7.25pt\noindent}}
\def\supertitle#1#2#3#4{\par\ifnum\prevgraf>0 \npageno=0 \fi
   \global\shortno=0\global\secno=0\global\tabno=0\global
   \figno=0\global\footno=0\global\refno=0\goodbreak\ifnum\npageno=0
   \par\vskip 1.3cm plus2.5mm minus2.5mm\fi\noindent{\textfont1=\mitc
   \rmc #1}\par
   \penalty1000\vskip1cm plus 3mm minus 7mm{\noindent\textfont1=\mitd\rmd
   #2$\vphantom{\hbox{\rmd y}}$}\par\penalty1000\vskip8pt\noindent{\itb
   #3$\vphantom{\hbox{\itb y}}$}{\par\penalty1000\vskip7.5pt\baselineskip
   11.5pt\hrule height 0.1pt\par\penalty1000\vskip7pt{\tenpoint\noindent\rm
   #4$\vphantom{y}$}\par\penalty1000\vskip6pt\hrule height 0.1pt\par
   \penalty1000\vskip9pt\noindent}}
\def\ssupertitle#1#2#3#4{\par\ifnum\prevgraf>0 \npageno=0 \fi
   \global\shortno=0\global\secno=0\global\tabno=0\global
   \figno=0\global\footno=0\global\refno=0\goodbreak\ifnum\npageno=0
   \par\vskip 1.1cm plus2.1mm minus2.1mm\fi\noindent{\textfont1=\mitb
   \rmb #1$\vphantom{\hbox{\rmb y}}$}\par
   \penalty1000\vskip8mm plus 3mm minus 6mm{\noindent\textfont1=\mitc\rmc
   #2$\vphantom{\hbox{\rmc y}}$}\par\penalty1000\vskip8pt\noindent{\ita
   #3$\vphantom{\hbox{\ita y}}$}{\par\penalty1000\vskip6pt\baselineskip
   10pt\hrule height
   0.1pt\par\penalty1000\vskip5.5pt{\ninepoint\noindent\rm
   #4$\vphantom{y}$}\par\penalty1000\vskip6pt\hrule height 0.1pt\par
   \penalty1000\vskip9pt\noindent}}
%
%
\def\firstreftitle#1#2#3#4{\par\ifnum\prevgraf>0 \npageno=0 \fi
   \global\shortno=0\global\secno=0\global\tabno=0\global
   \figno=0\global\footno=0\global\refno=0\goodbreak\ifnum\npageno=0
   \par\vskip 1.5cm plus2.5mm minus2.5mm\fi\noindent{\textfont1=\mitd
   \rmc Progress in Meteor Science}{\par\penalty1000
\vskip2mm plus 0.5mm minus 0.5mm\baselineskip11.5pt{\tenpoint\noindent\it
Articles in this section have been formally refereed by at least one
professional and one experienced, knowledgeable amateur meteor worker,
and deal with global analyses of meteor data, methods for meteor
observing and data reduction, observations with professional equipment, or
theoretical studies.}\par}\par\penalty1000
\vskip7.5mm plus 2.5mm minus 2.5mm{\noindent\textfont1=\mitd\rmd
#1$\vphantom{\hbox{\rmd y}}$}\par\penalty1000\vskip8pt\noindent{\itb
#2$\vphantom{\hbox{\itb y}}$}{\tenpoint\rm
\baselineskip11.5pt\footnote{}{\hskip-\parindent #3\smallnewpar\it
WGN, the Journal of the
International Meteor Organization, Vol.~\the\volno, No.~\the\nrno,
\month~\the\yearno, pp.~\the\firstpageno--\the\lastpageno.}}%
{\par\penalty1000\vskip7.5pt\baselineskip
11.5pt\hrule height 0.1pt\par\penalty1000\vskip7pt{\tenpoint\noindent\rm
#4$\vphantom{y}$}\par\penalty1000\vskip6pt\hrule height 0.1pt\par
\penalty1000\vskip9pt\noindent}}%
\def\reftitle#1#2#3#4{\par\ifnum\prevgraf>0 \npageno=0 \fi
   \global\shortno=0\global\secno=0\global\tabno=0\global
   \figno=0\global\footno=0\global\refno=0\goodbreak\ifnum\npageno=0
   \par\vskip 1.5cm plus2.5mm minus2.5mm\fi\noindent{\textfont1=\mitd\rmd
#1$\vphantom{\hbox{\rmd y}}$}\par\penalty1000\vskip8pt\noindent{\itb
#2$\vphantom{\hbox{\itb y}}$}{\tenpoint\rm
\baselineskip11.5pt\footnote{}{\hskip-\parindent #3\smallnewpar\it
WGN, the Journal of the
International Meteor Organization, Vol.~\the\volno, No.~\the\nrno,
\month~\the\yearno, pp.~\the\firstpageno--\the\lastpageno.}}%
{\par\penalty1000\vskip7.5pt\baselineskip
11.5pt\hrule height 0.1pt\par\penalty1000\vskip7pt{\tenpoint\noindent\rm
#4$\vphantom{y}$}\par\penalty1000\vskip6pt\hrule height 0.1pt\par
\penalty1000\vskip9pt\noindent}}%
\def\noabsupertitle#1#2#3{\par\ifnum\prevgraf>0 \npageno=0 \fi
   \global\shortno=0\global\secno=0\global\tabno=0\global
   \figno=0\global\footno=0\global\refno=0\goodbreak\ifnum\npageno=0
   \par\vskip 2cm plus2.5mm minus2.5mm\fi\noindent{\rmc #1}\par
   \penalty1000\vskip1cm plus 5mm minus 5mm{\noindent\rmd
   #2$\vphantom{\hbox{\rmd y}}$}\par\penalty1000\vskip8pt\noindent{\itb
   #3$\vphantom{\hbox{\itb y}}$}{\par\penalty1000\vskip7.5pt
   \hrule height 0.1pt\par\penalty1000\vskip9pt\noindent}}
\def\noabtitle#1#2{\par\ifnum\prevgraf>0 \npageno=0\fi
   \global\shortno=0\global\secno=0\global\tabno=0\global
   \figno=0\global\footno=0\global\refno=0\goodbreak\ifnum\npageno=0
   \par\vskip2cm plus 2.5mm minus 2.5mm\fi
   {\noindent\rmd #1$\vphantom{\hbox{\rme y}}$}\par\penalty1000\vskip8pt
   \noindent{\itb #2$\vphantom{\hbox{\itb y}}$}\par\penalty1000\vskip7.5pt
   \hrule height0.1pt\par\penalty1000\vskip9pt\noindent}
\def\snoabtitle#1#2{\par\ifnum\prevgraf>0 \npageno=0\fi
   \global\shortno=0\global\secno=0\global\tabno=0\global
   \figno=0\global\footno=0\global\refno=0\goodbreak\ifnum\npageno=0
   \par\vskip1.6cm plus 2mm minus 2mm\fi
   {\noindent\textfont1=\mitc
   \rmc #1$\vphantom{\hbox{\rmc y}}$}\par\penalty1000\vskip6.5pt
   \noindent{\ita #2$\vphantom{\hbox{\ita y}}$}\par\penalty1000\vskip6pt
   \hrule height0.1pt\par\penalty1000\vskip7.25pt\noindent}
\def\shorttitle#1#2{\global\secno=0\global\tabno=0\global
   \figno=0\global\footno=0\global\refno=0{\noindent
   \rmd #1$\vphantom{\hbox{\rme y}}$}\par
   \penalty1000\vskip8pt\noindent{\itb #2$\vphantom{\hbox{\itb y}}$}\par
   \penalty1000\vskip7.5pt\hrule height0.1pt\par\penalty1000
   \vskip9pt\noindent}
\def\sshorttitle#1#2{\global\secno=0\global\tabno=0\global
   \figno=0\global\footno=0\global\refno=0{\noindent
   \rmc #1$\vphantom{\hbox{\rmd y}}$}\par
   \penalty1000\vskip6.5pt\noindent{\ita #2$\vphantom{\hbox{\ita y}}$}\par
   \penalty1000\vskip6pt\hrule height0.1pt\par\penalty1000
   \vskip7.25pt\noindent}
\def\noauttitle#1{\global\secno=0\global\tabno=0\global\figno=0
 \global\footno=0\global\refno=0{\noindent\rmd #1$\vphantom{\hbox{\rmd y}}$}
   \par\vskip 7.5pt\hrule height
   0.1pt\par\penalty1000\vskip9pt\noindent}
\def\noabsemibigtitle#1#2#3#4{\global\shortno=0\global\secno=0
   \global\tabno=0\global\figno=0\global\footno=0\global\refno=0\goodbreak
   \ifnum\npageno=0 \par\vskip 2cm plus2.5mm  minus2.5mm\fi{\noindent
   \rmc #1$\vphantom{\hbox{\rmd y}}$}\par
   \penalty1000\vskip7pt\noindent{\rmd #2$\vphantom{\hbox{\rmc y}}$}\par
   \penalty1000\vskip7pt\noindent{\rmd #3$\vphantom{\hbox{\rmc y}}$}\par
 \par\penalty1000\vskip7pt\noindent{\itb #4$\vphantom{\hbox{\itb y}}$}{\par
   \penalty1000\vskip7.5pt\baselineskip11.5pt\hrule height 0.1pt\par
   \penalty1000\vskip9pt\noindent}}
\def\supersemititle#1#2#3#4#5{\global\secno=0\global\tabno=0\global\figno=0
   \global\footno=0\global\refno=0\goodbreak\ifnum\npageno=0
   \par\vskip 1.3cm plus2.5mm minus2.5mm\fi\noindent{\textfont1=\mitd
   \rmc #1$\vphantom{\hbox{\rmc y}}$}\par
   \penalty1000\vskip7pt\noindent{\textfont1=\mitd\rmd
   #2$\vphantom{\hbox{\rmd y}}$}\par
   \penalty1000\vskip7pt\noindent{\rmd #3$\vphantom{\hbox{\rmd y}}$}\par
 \par\penalty1000\vskip8pt\noindent{\itb #4$\vphantom{\hbox{\itb y}}$}{\par
   \penalty1000\vskip7.5pt\baselineskip11.5pt\hrule height 0.1pt\par
   \penalty1000\vskip7pt{\tenpoint\noindent\rm
   #5$\vphantom{y}$}\par\penalty1000\vskip6pt\hrule height 0.1pt\par
   \penalty1000\vskip9pt\noindent}}
\def\noabsupersemititle#1#2#3#4{\global\secno=0\global\tabno=0\global\figno=0
   \global\footno=0\global\refno=0\goodbreak\ifnum\npageno=0
   \par\vskip 1.3cm plus2.5mm minus2.5mm\fi\noindent{\textfont1=\mitd
   \rmc #1$\vphantom{\hbox{\rmc y}}$}\par
   \penalty1000\vskip7pt\noindent{\textfont1=\mitd\rmd
   #2$\vphantom{\hbox{\rmd y}}$}\par
   \penalty1000\vskip7pt\noindent{\rmd #3$\vphantom{\hbox{\rmd y}}$}\par
 \par\penalty1000\vskip8pt\noindent{\itb #4$\vphantom{\hbox{\itb y}}$}{\par
   \penalty1000\vskip7.5pt\baselineskip11.5pt\hrule height 0.1pt\par
   \penalty1000\vskip9pt\noindent}}
\def\ssupersemititle#1#2#3#4#5{\global\secno=0\global\tabno=0\global\figno=0
   \global\footno=0\global\refno=0\goodbreak\ifnum\npageno=0
   \par\vskip 1.3cm plus2.5mm minus2.5mm\fi\noindent{\textfont1=\mitb
   \rmb #1$\vphantom{\hbox{\rmb y}}$}\par
   \penalty1000\vskip6.5pt\noindent{\rmc #2$\vphantom{\hbox{\rmc y}}$}\par
   \penalty1000\vskip6.5pt\noindent{\rmb #3$\vphantom{\hbox{\rmb y}}$}\par
 \par\penalty1000\vskip8pt\noindent{\ita #4$\vphantom{\hbox{\ita y}}$}{\par
   \penalty1000\vskip7pt\baselineskip10.35pt\hrule height 0.1pt\par
   \penalty1000\vskip6pt{\ninepoint\noindent\rm
   #5$\vphantom{y}$}\par\penalty1000\vskip5.75pt\hrule height 0.1pt\par
   \penalty1000\vskip8pt\noindent}}

\def\bigtitle#1#2#3#4{\par\ifnum\prevgraf>0 \npageno=0 \fi
   \global\shortno=0\global\secno=0\global\tabno=0\global
   \figno=0\global\footno=0\global\refno=0\goodbreak\ifnum\npageno=0
   \par\vskip 2cm plus2.5mm  minus2.5mm\fi{\noindent\textfont1=\mitd\rmd #1$
   \vphantom{\hbox{\rmd y}}$}\par\penalty1000\vskip8pt\noindent
   {\textfont1=\mitd\rmd #2$\vphantom{\hbox{\rmd y}}$}
\par\penalty1000\vskip 8pt\noindent{\itb #3$\vphantom{\hbox{\itb y}}$}{\par
   \penalty1000\vskip7.5pt\baselineskip11.5pt\hrule height 0.1pt\par
   \penalty1000\vskip 7pt{\tenpoint\noindent\rm #4$\vphantom{y}$}\par
   \penalty1000\vskip6pt\hrule height 0.1pt\par\vskip9pt\noindent}}
\def\noabbigtitle#1#2#3{\global\shortno=0\global\secno=0\global\tabno=0\global
   \figno=0\global\footno=0\global\refno=0\goodbreak\ifnum\npageno=0
   \par\vskip 2cm plus2.5mm  minus2.5mm\fi{\noindent
   \rmd #1$\vphantom{\hbox{\rmd y}}$}\par
   \penalty1000\vskip8pt\noindent{\rmd #2$\vphantom{\hbox{\rmd y}}$}
 \par\penalty1000\vskip 8pt\noindent{\itb #3$\vphantom{\hbox{\itb y}}$}{\par
   \penalty1000\vskip7.5pt\hrule height 0.1pt\par
   \penalty1000\vskip9pt\noindent}}
\def\snoabbigtitle#1#2#3{\global\shortno=0
   \global\secno=0\global\tabno=0\global
   \figno=0\global\footno=0\global\refno=0\goodbreak\ifnum\npageno=0
   \par\vskip 1.6cm plus2mm  minus2mm\fi{\noindent\textfont1=\mitc%
   \rmc #1$\vphantom{\hbox{\rmc y}}$}\par
   \penalty1000\vskip6pt\noindent{\rmc #2$\vphantom{\hbox{\rmc y}}$}
 \par\penalty1000\vskip 6pt\noindent{\ita #3$\vphantom{\hbox{\ita y}}$}{\par
   \penalty1000\vskip6pt\hrule height 0.1pt\par
   \penalty1000\vskip7.25pt\noindent}}
\def\snoabsemititle#1#2#3{\global\shortno=0
   \global\secno=0\global\tabno=0\global
   \figno=0\global\footno=0\global\refno=0\goodbreak\ifnum\npageno=0
   \par\vskip 1.6cm plus2mm  minus2mm\fi{\noindent\textfont1=\mitc%
   \rmc #1$\vphantom{\hbox{\rmc y}}$}\par
   \penalty1000\vskip6pt\noindent{\rmb #2$\vphantom{\hbox{\rmb y}}$}
   \par\penalty1000\vskip 6pt\noindent{\ita #3$\vphantom{\hbox{\ita y}}$}{\par
   \penalty1000\vskip6pt\hrule height 0.1pt\par
   \penalty1000\vskip7.25pt\noindent}}
\def\snoabsupertitle#1#2#3{\global\shortno=0
   \global\secno=0\global\tabno=0\global
   \figno=0\global\footno=0\global\refno=0\goodbreak\ifnum\npageno=0
   \par\vskip 1.6cm plus2mm  minus2mm\fi{\noindent\textfont1=\mitb%
   \rmb #1$\vphantom{\hbox{\rmb y}}$}\par
   \penalty1000\vskip6pt\noindent{\rmc #2$\vphantom{\hbox{\rmc y}}$}
   \par\penalty1000\vskip 6pt\noindent{\ita #3$\vphantom{\hbox{\ita y}}$}{\par
   \penalty1000\vskip6pt\hrule height 0.1pt\par
   \penalty1000\vskip7.25pt\noindent}}
\def\snoabbigsemititle#1#2#3#4{\global\shortno=0
   \global\secno=0\global\tabno=0\global
   \figno=0\global\footno=0\global\refno=0\goodbreak\ifnum\npageno=0
   \par\vskip 1.6cm plus2mm  minus2mm\fi{\noindent
   \rmc #1$\vphantom{\hbox{\rmc y}}$}\par
   \penalty1000\vskip6pt{\noindent
   \rmc #2$\vphantom{\hbox{\rmc y}}$}\par
   \penalty1000\vskip6pt\noindent{\rmb #3$\vphantom{\hbox{\rmb y}}$}
 \par\penalty1000\vskip 6pt\noindent{\ita #4$\vphantom{\hbox{\ita y}}$}{\par
   \penalty1000\vskip6pt\hrule height 0.1pt\par
   \penalty1000\vskip7.25pt\noindent}}
\def\semititle#1#2#3#4{\global\shortno=0\global\secno=0\global\tabno=0\global
   \figno=0\global\footno=0\global\refno=0\goodbreak\ifnum\npageno=0
   \par\vskip 2cm plus2.5mm  minus2.5mm\fi{\noindent
   \rmd #1$\vphantom{\hbox{\rmd y}}$}\par
   \penalty1000\vskip7pt\noindent{\rmc #2$\vphantom{\hbox{\rmc y}}$}
 \par\penalty1000\vskip 7pt\noindent{\itb #3$\vphantom{\hbox{\itb y}}$}{\par
   \penalty1000\vskip7.5pt\baselineskip11.5pt\hrule height 0.1pt\par
   \penalty1000\vskip 7pt{\tenpoint\noindent\rm #4$\vphantom{y}$}\par
   \penalty1000\vskip6pt\hrule height 0.1pt\par\vskip9pt\noindent}}
\def\noabsemititle#1#2#3{\global\shortno=0\global\secno=0\global\tabno=0
    \global\figno=0\global\footno=0\global\refno=0\goodbreak
    \ifnum\npageno=0 \par\vskip 2cm plus2.5mm  minus2.5mm\fi{\noindent
    \rmd #1$\vphantom{\hbox{\rmd y}}$}\par
   \penalty1000\vskip7pt\noindent{\rmc #2$\vphantom{\hbox{\rmc y}}$}
 \par\penalty1000\vskip7pt\noindent{\itb #3$\vphantom{\hbox{\itb y}}$}{\par
   \penalty1000\vskip7.5pt\baselineskip11.5pt\hrule height 0.1pt\par
   \penalty1000\vskip9pt\noindent}}
\def\noautshortnote#1{\global\secno=0\global\tabno=0\global
   \figno=0\global\footno=0\global\refno=0\goodbreak\ifnum\npageno=0
   \ifnum\shortno=0 \vskip6mm plus 2mm minus 2mm\noindent
   \hrule height 1.2pt\par\vskip8.5mm plus 2.5mm minus 2.5mm
   \else\vskip17mm plus 4mm minus 3mm\fi\fi
   \global\shortno=1\noindent\noauttitle{#1}}
\def\shortnote#1#2{\global\secno=0\global\tabno=0\global
   \figno=0\global\footno=0\global\refno=0\goodbreak\ifnum\npageno=0
   \ifnum\shortno=0 \vskip6mm plus 2mm minus 2mm\noindent
   \hrule height 1.2pt\par\vskip8.5mm plus 2.5mm minus 2.5mm
   \else\vskip17mm plus 4mm minus 3mm\fi\fi
   \global\shortno=1\noindent\shorttitle{#1}{#2}}
\def\sshortnote#1#2{\global\secno=0\global\tabno=0\global
   \figno=0\global\footno=0\global\refno=0\goodbreak\ifnum\npageno=0
   \ifnum\shortno=0 \vskip5mm plus 1.75mm minus 1.75mm\noindent
   \hrule height 1.2pt\par\vskip7.25mm plus 2mm minus 2mm
   \else\vskip15.5mm plus 3.5mm minus 2.5mm\fi\fi
   \global\shortno=1\noindent\sshorttitle{#1}{#2}}
\def\shortnotes#1#2{\global\secno=0\global\tabno=0\global
   \figno=0\global\footno=0\global\refno=0\goodbreak\ifnum\npageno=0
   \ifnum\shortno=0 \vskip8.5mm plus 2.5mm minus 2.5mm
   \noindent\hrule height 1.2pt\par\vskip8.5mm plus 2.5mm minus 2.5mm
   \else\vskip17mm plus 5mm minus 5mm\fi\fi
 \global\shortno=1\noindent{\rmc Short Notes}\par\penalty10000\vskip8.5mm
    plus 2.5mm minus 2.5mm\noindent\shorttitle{#1}{#2}}
\parskip=0pt
\parindent=25pt
\def\newpar{\edef\next{\hangafter=\the\hangafter
    \hangindent=\the\hangindent}
    \par\penalty10000\ifnum\prevgraf>0 \global\npageno=0\fi
    \par\penalty-100\vskip 4pt plus 2pt minus 2pt\next%
    \edef\next{\ifnum\prevgraf>-\the\hangafter\prevgraf=0\hangafter=1
         \hangindent=0pt\else\prevgraf=\the\prevgraf\fi}
    \noindent\next}
\def\smallnewpar{\edef\next{\hangafter=\the\hangafter
    \hangindent=\the\hangindent}
    \par\penalty10000\ifnum\prevgraf>0 \global\npageno=0\fi
    \par\penalty-100\vskip 3pt plus 1.5pt minus 1pt\next%
    \edef\next{\ifnum\prevgraf>-\the\hangafter\prevgraf=0\hangafter=1
         \hangindent=0pt\else\prevgraf=\the\prevgraf\fi}
    \noindent\next}
\def\pensmallnewpar{\edef\next{\hangafter=\the\hangafter
    \hangindent=\the\hangindent}
    \par\penalty10000\ifnum\prevgraf>0 \global\npageno=0\fi
    \par\penalty10000\vskip 3pt plus 1.5pt minus 1pt\next%
    \edef\next{\ifnum\prevgraf>-\the\hangafter\prevgraf=0\hangafter=1
         \hangindent=0pt\else\prevgraf=\the\prevgraf\fi}
    \noindent\next}
\newcount\npageno
\newcount\shortno
\newcount\secno
\newcount\tabno
\newcount\figno
\newcount\refno
\newdimen\figindent
\newdimen\figheight
\newdimen\newfigheight
\newdimen\figwidth
\npageno=0
\shortno=0
\secno=0
\refno=0
\tabno=0
\figno=0
\figindent=0sp
\figheight=0sp
\newfigheight=0sp
\figwidth=0sp
\def\section#1{\par\global\npageno=0
    \global\advance\secno by 1\ifnum\prevgraf=0 \relax
    \else\penalty-100\vskip 12pt plus 6pt minus 4.5pt\fi
    {\textfont0=\twelvebf\textfont1=\twelvebi\textfont2=\bsi
     \scriptfont0=\ninebf
    \noindent{\bf\the\secno.\ #1}}\par\penalty10000
    \vskip 4pt plus 2pt minus2pt\noindent}%
\def\smallsection#1{\par\global\npageno=0
    \global\advance\secno by 1\ifnum\prevgraf=0 \relax
    \else\penalty-100\vskip 10pt plus 5pt minus 4pt\fi
    {\textfont0=\tenbf\textfont1=\tenbi\textfont2=\si
    \noindent{\bf\the\secno.\ #1}}\newpar}
\def\appsection#1{\par\global\npageno=0
   \ifnum\prevgraf=0 \relax
   \else\penalty-100\vskip 12pt plus 6pt minus 4.5pt\fi
   {\textfont0=\twelvebf\textfont1=\twelvebi\textfont2=\bsi
   \noindent{\bf#1}}\newpar}
\def\smallappsection#1{\par\global\npageno=0
   \ifnum\prevgraf=0 \relax
   \else\penalty-100\vskip 10pt plus 5pt minus 4pt\fi
   {\textfont0=\tenbf\textfont1=\tenbi\textfont2=\si
   \noindent{\bf#1}}\newpar}
\def\artref#1#2#3#4{\par\global\advance\refno by 1\ifnum\prevgraf=0 \relax
    \else\penalty-100\vskip 2pt plus 0.5pt minus 0.5pt\fi
    \hangindent\parindent\noindent\hbox to \parindent{[$
    \the\refno$]\hfill}#1, ``#2'', {\it #3\/}, #4.}
\def\sartref#1#2#3#4{\par\global\advance\refno by 1\ifnum\prevgraf=0 \relax
    \else\penalty-100\vskip 1.5pt plus 0.5pt minus 0.5pt\fi
    \hangindent\parindent\noindent\hbox to \parindent{[$
    \the\refno$]\hfill}#1, ``#2'', {\it #3\/}, #4.}
\def\noautref#1#2#3{\par\global\advance\refno by 1\ifnum\prevgraf=0 \relax
    \else\penalty-100\vskip 2pt plus 0.5pt minus 0.5pt\fi
    \hangindent\parindent\noindent\hbox to \parindent{[$
    \the\refno$]\hfill}``#1'', {\it #2\/}, #3.}
\def\noartref#1#2#3{\par\global\advance\refno by 1\ifnum\prevgraf=0 \relax
    \else\penalty-100\vskip 2pt plus 0.5pt minus 0.5pt\fi
    \hangindent\parindent\noindent\hbox to \parindent{[$
    \the\refno$]\hfill}#1, {\it #2\/}, #3.}
\def\snoartref#1#2#3{\par\global\advance\refno by 1\ifnum\prevgraf=0 \relax
    \else\penalty-100\vskip 1.5pt plus 0.5pt minus 0.5pt\fi
    \hangindent\parindent\noindent\hbox to \parindent{[$
    \the\refno$]\hfill}#1, {\it #2\/}, #3.}
\def\noartnoautref#1#2{\par\global\advance\refno by 1\ifnum\prevgraf=0 \relax
    \else\penalty-100\vskip 2pt plus 0.5pt minus 0.5pt\fi
    \hangindent\parindent\noindent\hbox to \parindent{[$
    \the\refno$]\hfill}{\it #1\/}, #2.}
\def\bookref#1#2#3{\par\global\advance\refno by 1\ifnum\prevgraf=0 \relax
    \else\penalty-100\vskip 2pt plus 0.5pt minus 0.5pt\fi
    \hangindent\parindent\noindent\hbox to \parindent{[$
    \the\refno$]\hfill}#1, ``#2'', #3.}
\def\sbookref#1#2#3{\par\global\advance\refno by 1\ifnum\prevgraf=0 \relax
    \else\penalty-100\vskip 1.5pt plus 0.5pt minus 0.5pt\fi
    \hangindent\parindent\noindent\hbox to \parindent{[$
    \the\refno$]\hfill}#1, ``#2'', #3.}
\def\noautbookref#1#2{\par\global\advance\refno by 1\ifnum\prevgraf=0 \relax
    \else\penalty-100\vskip 2pt plus 0.5pt minus 0.5pt\fi
    \hangindent\parindent\noindent\hbox to \parindent{[$
    \the\refno$]\hfill}``#1'', #2.}
\def\shortref#1#2{\par\global\advance\refno by 1\ifnum\prevgraf=0 \relax
    \else\penalty-100\vskip 2pt plus 0.5pt minus 0.5pt\fi
    \hangindent\parindent\noindent\hbox to \parindent{[$
    \the\refno$]\hfill}#1, ``#2''.}
\def\sourceref#1{\par\global\advance\refno by 1\ifnum\prevgraf=0 \relax
    \else\penalty-100\vskip 2pt plus 0.5pt minus 0.5pt\fi
    \hangindent\parindent\noindent\hbox to \parindent{[$
    \the\refno$]\hfill}{\it #1.}}
\def\perscom#1#2{\par\global\advance\refno by 1\ifnum\prevgraf=0 \relax
    \else\penalty-100\vskip 2pt plus 0.5pt minus 0.5pt\fi
    \hangindent\parindent\noindent\hbox to \parindent{[$
    \the\refno$]\hfill}#1, {\it personal communications\/}, #2.}
\def\libref#1#2{\par\hangindent\parindent\noindent
   \hbox to \parindent{$\bullet$\hfill}{\it #1}\pensmallnewpar{\rm #2}\newpar}

\newbox\tabbox
\def\table#1#2{\global\npageno=0\advance\tabno by 1
    \setbox\tabbox=\vbox{\tenpoint\tabskip=0pt
    \offinterlineskip\halign{#2}}
       $$\vbox{\tenpoint\baselineskip11.5pt\ialign{\hfill$##$\crcr
                  \vtop{\hsize=\wd\tabbox\hangindent1.65truecm\noindent%
                  \hbox to 1.65truecm{Table \the\tabno\ -- \hfill}#1}\crcr
                  \noalign{\vskip8pt}
                  \box\tabbox\crcr}}$$}
\def\nocaptable#1{\global\npageno=0\advance\tabno by 1
    \setbox\tabbox=\vbox{\tenpoint\tabskip=0pt
    \offinterlineskip\halign{#1}}
       $$\vbox{\tenpoint\baselineskip11.5pt\ialign{\hfill$##$\crcr
                  \box\tabbox\crcr}}$$
}\def\smalltable#1#2{\global\npageno=0\advance\tabno by 1
    \setbox\tabbox=\vbox{\ninepoint\tabskip=0pt
    \offinterlineskip\halign{#2}}
       $$\vbox{\ninepoint\baselineskip10.35pt\ialign{\hfill$##$\crcr
                  \vtop{\hsize=\wd\tabbox\hangindent1.5truecm\noindent%
      \hbox to 1.5truecm{\ninerm Table \the\tabno\ -- \hfill}#1}\crcr
                  \noalign{\vskip8pt}
                  \box\tabbox\crcr}}$$}
\def\nocapsmalltable#1{\global\npageno=0\advance\tabno by 1
    \setbox\tabbox=\vbox{\ninepoint\tabskip=0pt
    \offinterlineskip\halign{#1}}
       $$\vbox{\ninepoint\baselineskip10.35pt\ialign{\hfill$##$\crcr
                  \box\tabbox\crcr}}$$}
\def\bigtable#1#2{\global\npageno=0\advance\tabno by 1
    \setbox\tabbox=\vbox{\tabskip=0pt
    \offinterlineskip\halign{#2}}
       $$\vbox{\ialign{\hfill$##$\crcr
                  \vtop{\hsize=\wd\tabbox\hangindent2truecm\noindent%
                  \hbox to 2truecm{Table \the\tabno\ -- \hfill}#1}\crcr
                  \noalign{\vskip9.5pt}
                  \box\tabbox\crcr}}$$}
\newbox\figbox
\newcount\hgtno
\newcount\basno
\def\figure#1#2#3#4#5{\par\global\advance\figno by 1 \ifnum\prevgraf=0
    \noindent
    \else\newpar\fi \figindent=#1\advance\figindent by 5mm%
    \setbox\figbox=\vbox{\tenpoint\baselineskip11.5pt\hsize
        =#1\hangindent1.75truecm\noindent\hbox to 1.75truecm{Figure \the\figno
        \ -- \hfill}#5}%
    \figheight=#2\advance\figheight by 16pt\advance\figheight by \ht\figbox
    \hgtno=\figheight \basno=\baselineskip \divide\hgtno by \basno
    \advance\hgtno by 1 \advance\hgtno by #3 \advance\figheight by #4%
    \hangindent=\the\figindent\hangafter=-\the\hgtno\noindent
    \hskip-\the\figindent$\smash{\hbox to \the\figindent{\vtop to
    \the\figheight{\hsize=#1\vfill\vskip8pt\vbox{\tenpoint
    \baselineskip11.5pt\hsize=#1\hangindent1.75truecm\noindent
    \hbox to 1.75truecm{Figure \the\figno\ --
    \hfill}#5}\vskip8pt\ }\hfill}}$}
\def\smallboxfigure#1#2#3#4#5{\par\global\advance\figno by 1
\ifnum\prevgraf=0
    \noindent
    \else\smallnewpar\fi \figindent=#1\advance\figindent by 5mm%
    \figwidth=#1\advance\figwidth by -2.4pt
    \setbox\figbox=\vbox{\ninepoint\baselineskip10.35pt\rm\hsize
     =#1\hangindent1.65truecm\noindent\hbox to 1.65truecm{Figure \the\figno
        \ -- \hfill}#5}%
    \figheight=#2\advance\figheight by 16pt\advance\figheight by \ht\figbox
    \hgtno=\figheight \basno=\baselineskip \divide\hgtno by \basno
    \advance\hgtno by 1 \advance\hgtno by #3 \advance\figheight by #4%
    \hangindent=\the\figindent\hangafter=-\the\hgtno
    \newfigheight=#2\advance\newfigheight by -2.4pt
    \noindent \hskip-\the\figindent$\smash{\hbox to \the\figindent{\vtop to
    \the\figheight{\vfill\hsize=#1\vskip-3pt\ialign to
    #1{\vrule##width1.2pt&\hfill\hbox to
    \figwidth{\hfil##\hfil}\hfill&\vrule##width1.2pt\crcr
    \noalign{\hrule height1.2pt}%
    height\newfigheight&\ &height\newfigheight\crcr
    \noalign{\hrule height1.2pt}}\vfill\vskip8pt\vbox{
   \ninepoint\baselineskip10.35pt\rm\hsize=#1\hangindent1.65truecm\noindent
    \hbox to 1.65truecm{Figure \the\figno\ --\hfill}#5}\vskip0pt\
    }\hfill}}$}
\def\nocapsmallfigure#1#2#3#4{\par\global\advance\figno by 1 \ifnum\prevgraf=0
    \noindent
    \else\smallnewpar\fi \figindent=#1\advance\figindent by 5mm%
    \figheight=#2\advance\figheight by 8pt
    \hgtno=\figheight \basno=\baselineskip \divide\hgtno by \basno
    \advance\hgtno by 1 \advance\hgtno by #3 \advance\figheight by #4%
    \hangindent=\the\figindent\hangafter=-\the\hgtno\noindent
    \hskip-\the\figindent$\smash{\hbox to \the\figindent{\vtop to
    \the\figheight{\hsize=#1\vfill}\hfill}}$}
\def\fullfigure#1#2#3#4{\par\global\npageno=0
    \global\advance\figno by 1 \ifnum\prevgraf=0 \noindent
    \else\newpar\fi
    \setbox\figbox=\vbox{\tenpoint\baselineskip11.5pt\hsize
    =#1\hangindent1.75truecm\noindent\hbox to 1.75truecm{Figure \the\figno
        \ -- \hfill}#3}%
    \figheight=#2\advance\figheight by 16pt\advance\figheight by \ht\figbox
    $$\vbox to \figheight{\hsize=#1\vfill#4\vskip8pt\vbox{
     \tenpoint\baselineskip11.5pt\hsize=#1\hangindent1.75truecm\noindent
     \hbox to 1.75truecm{Figure \the\figno\ --\hfill}#3}\vskip2.5pt}$$}
\def\plaktable#1#2#3{\par\global\npageno=0
    \global\advance\tabno by 1 \ifnum\prevgraf=0 \noindent
    \else\newpar\fi
    \setbox\figbox=\vbox{\tenpoint\baselineskip11.5pt\hsize
    =#1\hangindent1.75truecm\noindent\hbox to 1.75truecm{Table
    \the\tabno\ -- \hfill}#3}%
    \figheight=#2\advance\figheight by 16pt\advance\figheight by \ht\figbox
    $$\vbox to \figheight{\hsize=#1\vbox{\tenpoint\baselineskip11.5pt
     \hsize=#1\hangindent1.75truecm\noindent
     \hbox to 1.75truecm{Table \the\tabno\ --\hfill}#3}\vfill}$$}
\def\fullboxfigure#1#2#3{\par\global\npageno=0
    \global\advance\figno by 1 \ifnum\prevgraf=0 \noindent
    \else\newpar\fi
\figwidth=#1\advance\figwidth by -2.4pt
    \setbox\figbox=\vbox{\tenpoint\baselineskip11.5pt\hsize
    =#1\hangindent1.75truecm\noindent\hbox to 1.75truecm{Figure \the\figno
        \ -- \hfill}#3}%
    \figheight=#2\advance\figheight by 24pt\advance\figheight by
\ht\figbox
\newfigheight=#2\advance\newfigheight by -2.4pt
    $$\vbox to
\figheight{\vfill\vskip8pt\ialign to #1{\vrule##width1.2pt&\hfill\hbox to \figwidth{\hfil##\hfil}\hfill&\vrule##width1.2pt\crcr
\noalign{\hrule height1.2pt}
height\newfigheight&\ &height\newfigheight\crcr
\noalign{\hrule height1.2pt}}\vfill\vskip8pt\vbox{
     \tenpoint\baselineskip11.5pt\hsize=#1\hangindent1.75truecm\noindent
     \hbox to 1.75truecm{Figure \the\figno\
     --\hfill}#3}\vskip2.5pt}$$}
\def\smallfullboxfigure#1#2#3{\par\global\npageno=0
    \global\advance\figno by 1 \ifnum\prevgraf=0 \noindent
    \else\newpar\fi
\figwidth=#1\advance\figwidth by -2.4pt
    \setbox\figbox=\vbox{\ninepoint\baselineskip10.35pt\hsize
    =#1\hangindent1.65truecm\noindent\hbox to 1.65truecm{Figure \the\figno
        \ -- \hfill}#3}%
    \figheight=#2\advance\figheight by 24pt\advance\figheight by
\ht\figbox
\newfigheight=#2\advance\newfigheight by -2.4pt
    $$\vbox to
\figheight{\vfill\vskip8pt\ialign to
#1{\vrule##width1.2pt&\hfill\hbox to
\figwidth{\hfil##\hfil}\hfill&\vrule##width1.2pt\crcr
\noalign{\hrule height1.2pt}
height\newfigheight&\ &height\newfigheight\crcr
\noalign{\hrule height1.2pt}}\vfill\vskip8pt\vbox{
     \ninepoint\baselineskip10.35pt\hsize=#1\hangindent1.65truecm\noindent
     \hbox to 1.65truecm{Figure \the\figno\ --\hfill}#3}\vskip2.5pt}$$}
\def\smallfullfigure#1#2#3{\par\global\npageno=0
    \global\advance\figno by 1 \ifnum\prevgraf=0 \noindent
    \else\newpar\fi
    \setbox\figbox=\vbox{\ninepoint\baselineskip10.35pt\hsize
     =#1\hangindent1.65truecm\noindent\hbox to 1.65truecm{Figure \the\figno
        \ -- \hfill}#3}%
    \figheight=#2\advance\figheight by 16pt\advance\figheight by \ht\figbox
    $$\vbox to \figheight{\hsize=#1\vfill\vskip8pt\vbox{
     \ninepoint\baselineskip10.35pt\hsize=#1\hangindent1.65truecm\noindent
     \hbox to 1.65truecm{Figure \the\figno\ --\hfill}#3}\vskip2.5pt}$$}
\def\nocapsmallfullfigure#1#2#3{\par\global\npageno=0
    \global\advance\figno by 1 \ifnum\prevgraf=0 \noindent
    \else\newpar\fi
    \setbox\figbox=\vbox{\ninepoint\baselineskip10.35pt\hsize
     =#1{#3}}%
    \figheight=#2\advance\figheight by 16pt\advance\figheight by \ht\figbox
    $$\vbox to \figheight{\hsize=#1\vfill\vskip8pt\vbox{
     \ninepoint\baselineskip10.35pt\hsize=#1{\noindent #3}}\vskip2.5pt}$$}

\newskip\basskip
\basskip=11.5pt
\def\racco#1{\hbox to 10pt{$\smash{\vcenter{\hbox{$\left.\vphantom{\vcenter{
\vrule height #1\basskip}}\right\}$}}\hfill}$}}
\def\raccol#1{$\smash{\raise 0.5\basskip\hbox{#1}}$}

\def\t{\hbox{\rm --}}

\def\0{\phantom{0}}
\def\1{\phantom{.0}}
\def\E{\hbox to 6truemm{\hfill E\hfill}}
\def\W{\hbox to 6truemm{\hfill W\hfill}}
\def\N{\hbox to 6truemm{\hfill N\hfill}}
\def\S{\hbox to 6truemm{\hfill S\hfill}}
\mathchardef\g="020E
\mathchardef\h="0068
\mathchardef\m="006D
\mathchardef\s="0073
\newbox\help
\newbox\punt
\newskip\terug
\newskip\vooruit
\def\dg{{}^\g\setbox\help=\hbox{${}^\g$}\setbox\punt=\hbox{$.$}\skip
\terug=-.5\wd\help plus0pt minus0pt\advance\skip\terug by -0.17em plus0em
minus0em\hskip\skip\terug\skip\vooruit=-\skip\terug\advance\skip\vooruit by
-\wd\punt.\hskip\skip\vooruit}
\def\dh{{}^\h\setbox\help=\hbox{${}^\h$}\setbox\punt=\hbox{$.$}\skip
\terug=-.5\wd\help plus0pt minus0pt\advance\skip\terug by -0.17em plus0em
minus0em\hskip\skip\terug\skip\vooruit=-\skip\terug\advance\skip\vooruit by
-\wd\punt.\hskip\skip\vooruit}
\def\dm{{}^\m\setbox\help=\hbox{${}^\m$}\setbox\punt=\hbox{$.$}\skip
\terug=-.5\wd\help plus0pt minus0pt\advance\skip\terug by -0.17em plus0em
minus0em\hskip\skip\terug\skip\vooruit=-\skip\terug\advance\skip\vooruit by
-\wd\punt.\hskip\skip\vooruit}
\def\ds{{}^\s\setbox\help=\hbox{${}^\s$}\setbox\punt=\hbox{$.$}\skip
\terug=-.5\wd\help plus0pt minus0pt\advance\skip\terug by -0.17em plus0em
minus0em\hskip\skip\terug\skip\vooruit=-\skip\terug\advance\skip\vooruit by
-\wd\punt.\hskip\skip\vooruit}
\def\dpr{{}^\prime\setbox\help=\hbox{${}^\prime$}\setbox\punt=\hbox{$.$}\skip
\terug=-.5\wd\help plus0pt minus0pt\advance\skip\terug by -0.17em plus0em
minus0em\hskip\skip\terug\skip\vooruit=-\skip\terug\advance\skip\vooruit by
-\wd\punt.\hskip\skip\vooruit}
\def\ddp{{}^{\prime\prime}\setbox\help=\hbox{${}^{\prime\prime}$}
\setbox\punt=\hbox{$.$}\skip
\terug=-.5\wd\help plus0pt minus0pt\advance\skip\terug by -0.17em plus0em
minus0em\hskip\skip\terug\skip\vooruit=-\skip\terug\advance\skip\vooruit by
-\wd\punt.\hskip\skip\vooruit}
\def\?{\hbox{\rm ?}}%
\def\advancepageno{\global\npageno=1\global\advance\pageno by 1\ifodd\pageno
    \global\hoffset=-1truecm\else\global\hoffset=0truecm\fi}
\ifodd\pageno\global\hoffset=0truecm\else\global\hoffset=-1truecm\fi
\def\makeheadline{\vbox to 0pt{\vskip-1.3truecm
     \line{\vbox to 9.75pt{}\the\headline}\vss}\nointerlineskip}
\newcount\volno
\newcount\nrno
\newcount\yearno
\headline={\elfpoint\sl\ifodd\pageno WGN, the Journal of the IMO
\the\volno:\the\nrno\ (\the\yearno)\hfill$\the\pageno$\else
$\the\pageno$\hfill WGN, the Journal of the IMO
\the\volno:\the\nrno\ (\the\yearno)\fi}
\newcount\footno
\global\footno=0
\def\nfoot#1{\global\npageno=0\advance\footno by 1
    {\tenpoint\baselineskip11.5pt\footnote{$^{\the\footno}$}{#1}}}
\def\smallnfoot#1{\global\npageno=0\advance\footno by 1%
{\ninepoint\baselineskip10.35pt\footnote{$^\the\footno$}{#1}}}%
\def\boxit#1{\vtop{\hrule height1.2pt \hbox{\vrule width1.2pt\kern3.5pt
    \vbox{\kern3.5pt#1\kern3.5pt}
    \kern3.5pt\vrule width1.2pt}\hrule height1.2pt}}
\hsize=17.2truecm
\hfuzz=2pt
\vfuzz=2pt
\vsize=25.2truecm
\voffset=-0.5truecm
\newcount\firstpageno
\newcount\lastpageno
\firstpageno=\pageno
\twelvepoint
\baselineskip 13.75pt
\nopagenumbers
\font\smallcaps=cmcsc12
\title{The Observation of Lunar Impacts}
{Costantino Sigismondi and Giovanni Imponente}%
{The intense activity of cratering on the Moon and in the inner
regions of the solar system was accomplished during the first $10^9$
years~[1]. Occasionally, some impact events occur even nowadays. In
Section~1, we treat, from a historical point of view, the
Earth-based observation of lunar impacts. In Section~2, we consider
the visibility conditions of such events evaluating the luminosity of
the background upon which an impact shines. In Section~3, the
luminosity of an impact is discussed. The occurrence of lunar
impact events outside of meteor shower periods is calculated using the
hourly rate of the sporadic meteors and their population
index. The evidence of a larger rate of impacts of meteoroids in the
past under these hypotheses is presentend in the last section.}%
{\tenpoint\rm\baselineskip11.5pt\footnote{}{\hskip-\parindent 
The authors are affiliated with the Department of Physics,
University of Rome ``La Sapienza,'' and ICRA, International Center for
Relativistic Astrophysics, P$^{\rm le}$ A.~Moro 2, I-00185 Rome,
Italy. The first author can be contacted at {\tt sigismondi@icra.it}.}}%
\section{Evidence of lunar impacts}%
The first important evidence of lunar impacts has been the observation
recorded on June 18, 1178 (Julian calendar), by a few men after sunset
and registered in the chronicles of Gervase of Canterbury [2]. They
observed the upper horn of a crescent moon to have split with fire and
sparks emanating from the division point. This report was interpreted
to be a description of events related to the formation of the lunar
crater Giordano Bruno [3].
\newpar
On November~18 and~19, 1999, several impact events of initial
magnitude between~$+3$ and~$+7$ were recorded with a video tape and by
naked-eye near the center of the Moon's dark limb. Those flashes
resulted from Leonid impacts on the Moon, because the center of the
1899 dust trail would have passed $0.0002$ AU, from the selenocenter
around $4^\h 49^\m$ UT [4].
\newpar
The Giordano Bruno crater event is reliably exceptional, while the
last are more frequent. The observation from Padua, Italy, of a
probable lunar impac during the  
total eclipse of the Moon on January 21, 2000, done by C.~Sigismondi,
confirms that assertion.
\section{Luminosity of a lunar impact}%
Except for the very rare events such as the one reported by Gervase,
it is necessary to have a dark background upon which the flashes of
lunar meteors can be visible. Hence, there are three favorable
conditions for observing such a phenomenon:
\smallnewpar
\item{1.}in the ash-grey light of the Moon, close to New Moon;
\smallnewpar
\item{2.}far from the dark limb around First or Last Quarter;
\smallnewpar
\item{3.}during a total eclipse.
\smallnewpar
The last conditon is the most favorable one, but also the least frequent one. 
\newpar
In the case of the lunar Leonids the Moon phase was $62\%$, while for
the Giordano Bruno event, it was $11\%$ [5]
\newpar
{\it Ash-grey light}%
\smallnewpar
The maximum brightness of the Moon due to the ash-grey light occurs
during a total eclipse of the Sun. That value has been measured in
order to study the coronal aureola phenomenon [6], and it is
$10^{+9.3}$ fainter than the solar disk. It means that the Moon in
ash-grey light shines like a magnitude~$-3$ star, say its surface
brightness $B$ during the total eclipse of September 22, 1968, was
equal to  about  $+13$ per square arcsecond.
\newpar
{\it Total eclipse of the Moon}%
\smallnewpar
It is well known that a total lunar eclipse is measurable by
de-focusing a star and comparing its image with the eclipsed
Moon. Typically, the eclipsed Moon shines as a magnitude~0 star, but
exceptionally it shines as a magnitude~$+3$ star [7]. Hence, for the
surface brightness $B$, we typically find  $+16$ per square arcsecond.
\newpar
{\it The First or Last Quarter of the Moon}
\smallnewpar
The second case is intermediate with respect to the previous ones.
The dark quarter of the Moon is illuminated by a quarter of the
Earth, and therefore the amount of light coming from the Earth is
about half that of the New Moon case. It corresponds to a surface
brightness of about $+14$ per square arcsecond.
\newpar
We emphasize that all those surface brightnesses are calculated for the
{\it dark\/} area of the Moon. For naked-eye observations, however, if
the bright part of the Moon is not artificially occultated, because of
the atmospheric glare and the bleaching of the eye, the effective
luminosity background is even brighter than that of the ash-grey light. 
\section{The number of lunar meteors calculated by the sporadic rate}%
We assume that the 5 lunar events recorded by videotape during 1999
Leonids shower belong to the same ensemble of meteors with equal
population index $r=2.5$ and similar normalization factor such that the
${\rm ZHR}\approx 4000$.
\newpar
Analogously to the formula that gives the number of observed meteors
versus the limiting magnitude, already corrected by the position of
the radiant, we have 
$$N_{\rm obs}={\rm ZHR}\times r^{-(6.5-{\rm lm})}\times \Delta t_{\rm
h},$$
where $r$ is the population index, and we can affirm that the total
amount of the lunar Leonids observed around the peak time should be
very bright fireballs if they fall onto the Earth. Therefore, the
following equation, integrated over 2 hours of observation, is valid:
$$N_{\rm obs}=2\times {\rm ZHR} \times r^{-x}.$$
For $N_{\rm obs}=5$ in two hours of registration, $ZHR\approx 4000$,
and for the Leonids' population index $r=2.5$ we obtain $x\sim 8$. 
\newpar
To calculate the probability to have similar events out of the maxima
of meteors shower, one needs to consider the hourly rate HR and the
population index $r_{\rm Spor}$ of the sporadic meteors: ${\rm HR}\approx 10$
and $r_{\rm Spor}=3.4$ [8]. This yields the equation
$$N_{\rm exp}=10 \times 3.4^{-x} \times \Delta t_{\rm h}.$$
With $x\approx 8$, the expected value of sporadic lunar impacts to be
observed in two hours is $N_{\rm exp}\approx 1/900$. 
\newpar
In this calculation, we have considered the background conditions
similar to that ones of the lunar Leonids of 1999, with the Moon phase
at $62\%$. If we consider the illumination conditions during a Moon
eclipse, we can reach a background which is $2$ to $3$ magnitudes per square
arcsecond fainter. That fact implies an improvement of the
detection of the lunar meteor of a factor given by $r_{\rm Spor}^2$ to
$r_{\rm Spor}^3$, which is in the range 10--30. 
\newpar
Consequently, during the totality of a Moon eclipse (about two hours)
occurring outside one of meteors shower maxima, $N_{ecl}=1/90\t1/30$
lunar impacts are expected to be visible up to magnitude~$+8$. 
\newpar
This number should be reduced significantly if we assume a
non-homogeneous distribution of the matter inside the meteor stream at
which we refer our calculation. Since the Moon approached the core of
the 1899 dust trail much closer than the Earth, this is a real possibility.
As a limiting case, we consider the effective  ${\rm ZHR}=150\,000$ at
0.0002 AU apart of the center of 1899 dust trail (as in the 1966
Leonids shower [9]). For this limiting case, all computed
probabilities should be divided by 30--40.
\section{Kinetic energy assessment}%
We consider the temperature-kinetic energy relation for an object at
given impact velocity $v=41$~km/s, obtained averaging the
geocentric velocities of all known meteors showers.  
We evaluate the luminosity due to the radiation of the kinetic energy. 
The kinetic energy in calories ($C_M$) can be obtained by dividing the
kinetic energy $Mv^2/2$ computed in Joules by 4.18.
Such an amount of calories corresponds to an increment of temperature
(neglecting the melting heats) of
$\Delta T \approx C_M/M$ if $M$ is expressed in grams in the latter equation.
Combining both equations, we see that the mass cancels out, and we find
$\Delta T \approx 2\times10^5$~K. 
\newpar
In what follows, we consider a 10~g ice meteoroid.
Applying the Stefan-Boltzmann law of black-body emission at a given
temperature, we receive from a mass at $2\times 10^5$~K on the Moon
the following amount of radiation:
$$W_M= {\sigma T^4 \times A_M\over 4\pi\,d_{\rm Moon}^2} \approx 3
\times 10^{-8}\ \hbox{\rm W/m$^2$},$$
where $A_M \approx (M/\rho)^{2/3}$ is the area of the incoming
meteoroid (yielding $A_M\approx 5\ {\rm cm}^2$ in our case), and $d_{\rm
Moon}\approx 3.84 \times 10^8$~m is the Earth-Moon average distance.
\newpar
To calculate the visual magnitude, we must take into account that the
eye is sensitive in a range of wavelengths far from the peak
for a temperature of $T\approx2\times10^5$~K, which means a
reduction of a factor 10 for the effective temperature visible (say
$20\,000$~K), corresponding to a reduction of $10^4$ in the detected
intensity. Therefore, the energy flux of such an event of mass
$M=10$~g is $W_M\approx3\times 10^{-12}$~W/m$^2$, i.e., a magnitude 
$$m=-2.5\,\log{3\times 10^{-12}\over 3.7\times 10^{-9}}=7.7,$$
where $3.7\times 10^{-9}$~W/m$^2$ is the visual energy flux
corresponding to a magnitude~0 event [10]. The final value of the above
equation corresponds to the observed magnitudes of lunar Leonids~[4]. 
\newpar
Given the formula [11]
$$m_E=40-2.5\log(2.732\times 10^{10}M^{0.92}v_G^{3.91}),$$
where $m_E$ is the magnitude, M the meteoroid mass in grams, and $v_G$
its geocentric velocity in km/s, we can relate the mass $M$ of a lunar
meteoroid to its magnitude $m_E$ if it would fall in the Earth's
atmosphere. For a meteor with $v_G=41$~km/s and $M=10$~g, we have $m_E=-4.2$.
\newpar
Using the formula giving the number of meteors brighter than the
limiting magnitude lm, $N_{\rm obs}={\rm ZHR}\times r^{-(6.5-{\rm
lm})}$, we can obtain the expected hourly number of meteoroids
larger than mass $M$ by substituting $m_E$ from the previous equation
for ``lm.'' We assume moreover that such number is the same for the Moon.
(This last assumption neglects all the geometrical effects due to the
perspective of the line of sight. It is like considering that we can
observe only the events occurring on the central zone of the Moon
disk, and here we assume the same hourly rate observed from the
Earth.)
\newpar
For the 1999 Leonids, we expect $N_{\rm exp}= 4000\times
2.5^{-10.7}\approx 0.2$ per hour. Assuming ${\rm ZHR}=150\,000$,
because the Moon passed closer to the central zone of the meteor
stream $N_{\rm exp}\approx 8$ per hour, in agreement with the
observations. 
\newpar
For the sporadic rate we expect a much smaller number, namely
$N_{\rm exp}= 10 \times 3.4^{-10.7} \approx 2.1\times 10^{-5}$ per hour. 
During a total eclipse of the Moon of 2 hours, we expect only  $N_{\rm
exp} \approx 4.1 \times 10^{-5}$ events, say $N_{\rm exp} \approx 1/24000$.  
This value is only one order of magnitude smaller than the one derived
in Section~3.    
\section{History of cratering in the inner solar system}
Assuming that the impact rate of meteoroids has remained constant
during the last 5 billion years, we can estimate the expected value
of craters larger than diameter $D$ on the Moon's surface. 
We compare this number with the observational evidence.
\newpar
A crater of diameter $D \approx 4$~km---which corresponds to 2 arc seconds
at the Moon's distance---and depth $a\approx 100$~m is 
obtained from the impact of a meteoroid of mass $M$ and velocity
$v_G$. For simplicity, we assume that the kinetic energy of the
incoming body is almost able to raise the material contained in the
the crater to a height of $4D$ (for the Copernicus crater,
surrounded by a great radial structure visible at Full Moon, this
value is largely underestimated).  Hence,
$$Mv_G^2/2 = g_{\rm Moon} M_{\rm Crater}\times 4D,$$
where $g_{\rm Moon}$ is the gravitational acceleration for the Moon
($g_{\rm Moon}\approx 1.42$~m/s$^2$), and $M_{\rm Crater}$ is the mass
of the matter removed to form the crater. We find
$$M={2g_{\rm Moon}\rho_{\rm Moon}\pi a(D/2)^2\times 4D\over v_G^2}\,,$$
and, assuming again $v_G=41$~km/s and $\rho_{\rm Moon}\approx
2$~g/cm$^3$ (the lunar density), we finally obtain $M\approx 7.4
\times 10^7$~kg.  
According to the formula in [11], such a meteoroid falling on the Earth
should have a magnitude $m_E =-20$. The number of sporadic meteors
brighter than $m_E =-20$ scaled for the Moon's surface 
$A_{\rm Moon}/A_{\rm obs}$ gives us the expected number of craters
larger than 4 km above the Moon's surface. Here,
$A_{\rm obs}$ is the area from which an observer can see a meteor
trail occurring at an altitude of $h=80$~km in the Earth's atmosphere:
$$A_{\rm obs}\sim 2\pi R_{\rm Earth} h$$
($R_{\rm Earth}=6378$~km is the Earth's radius and $R_{\rm
Moon}=1739$~km is the Moon's radius).
Hence, we find
$${A_{\rm Moon}\over A_{\rm obs}}= {R^2_{\rm Moon}\over R_{\rm
Earth} h}=5.93.$$
Integration with the above scaling factor over $5\times 10^9$ years,
which is about $4.4\times 10^{13}$ hours, yields
$$N_{\rm exp}=5.93\times 4.4\cdot 10^{13}\times 10\times
(3.4)^{-26.5}\approx20.$$ 
This expected number is lower than the number of crater visible even
with an amateur equipment, and it demonstrates that in the past the
impact rate was larger than today. If we compare with the extrapolated
number of craters without considering the outflows of basalt of the
maria, the evidence becomes even more striking. 
\appsection{References}%
\bookref{J.K.~Beatty, C.C.~Petersen, A.~Chaikin, eds.}{The New Solar
System}{Fourth Edition, Cambridge Univ.\ Press, 1999}%
\noartref{J.B.~Hartung}{Journal of Geophysical Research\/\rm\
98:E5}{1993, p~.9141}%
\noartref{J.B.~Hartung}{Meteoritics\/\rm\ 11}{1976, pp.~187--194}%
\noartref{D.W.~Dunham}{{\rm in} IAU Circular\/\rm\ 7320}{1999}%
\bookref{E.C.~Downey}{{\smallcaps Ephem}---An Interactive Astronomical
Ephemeris Program}{Version 4.27 V, March 11, 1992; Copyright 1990,
1991; VGA Watch plots by J.D.~McDonald (free of charge)}%
\noartref{O.~Koutchmy, S.~Koutchmy}{Astron.\ Astrophys.\ Suppl.\/\rm\
13}{1974, pp.~295}%
\noartref{S.J.~O'Meara}{Sky and Telescope\/\rm\ 99:1}{2000, p.~109}%
\noartref{P.~Jenniskens}{Astron.\ Astrophys.\/\rm\ 287}{1994, p.~990}%
\noartref{Z.~Wu, I.P.~Williams}{Mon.\ Not.\ R.\ Astron.\ Soc.\/\rm\
280}{1996, p.~1210}%
\bookref{C.~Barbieri}{Lezioni di Astronomia}{Zanichelli, Bologna, 1999, p.~240}%
\noartref{R.~Arlt, P.~Brown}{WGN\/\rm\ 27:6}{December 1999, p.~278}%
\par\eject\end